\documentclass[12pt]{article}
\usepackage[dvips]{epsfig}

\begin{document}
\setlength{\unitlength}{1mm}

\def\lrD{\mathrel{{\cal D}\kern-1.em\raise1.75ex\hbox{$\leftrightarrow$}}}
\def\lr #1{\mathrel{#1\kern-1.25em\raise1.75ex\hbox{$\leftrightarrow$}}}

\newcommand{\al} \alpha
\newcommand{\la} \lambda
\newcommand{\om} \omega   \newcommand{\Om} \Omega
\newcommand{\bt} \beta
\newcommand{\p} \prime

\newcommand{\ba} {\begin{eqnarray}}
\newcommand{\ea} {\end{eqnarray}}

\newcommand{\be}{\begin{equation}}
\newcommand{\ee}{\end{equation}}
\newcommand{\n}[1]{\label{#1}}
\newcommand{\eq}[1]{Eq.(\ref{#1})}
\newcommand{\ind}[1]{\mbox{\tiny{#1}}}
\renewcommand\theequation{\thesection.\arabic{equation}}
\newcommand{\bk}[1] {\mbox{$ \langle #1 \rangle $}}

\title{Metric Fluctuation Corrections to  Hawking Radiation}
\author{\\
C. Barrab\`{e}s\thanks{e-mail: barrabes@celfi.phys.univ-tours.fr}  ${}^{1}$,
V. Frolov\thanks{e-mail: frolov@phys.ualberta.ca}
${}^{2}$ and 
R. Parentani\thanks{e-mail: 
parenta@celfi.phys.univ-tours.fr}
${}^{1}$ 
\date{}}
\maketitle
\noindent 
{
\\ $^{1}${\em Laboratoire de Math\'{e}matiques et Physique
Th\'{e}orique, 
CNRS UPRES A 6083,\\${ }\quad$
Universit\'{e} de Tours, 
37200 Tours, France}\\
$^{2}${ \em
Theoretical Physics Institute, Department of Physics, \ University of
Alberta, \\ ${ }\quad$ Edmonton, Canada T6G 2J1}
}
\bigskip

\begin{abstract}
We study  how fluctuations of the  black hole geometry affect the
properties of Hawking radiation.  Even though we treat the
fluctuations classically, we believe that the results so obtained
indicate what might be the  effects induced by quantum
fluctuations in a self consistent treatment. To
characterize the fluctuations,  we use the model introduced by York
in which  they  are described by an advanced Vaidya metric with a
fluctuating mass.  Under the assumption of spherical symmetry, we
solve the equation of null outgoing rays.
Then, by neglecting the greybody factor, we calculate
the late time corrections to the s-wave contributions
of the energy flux and the asymptotic
spectrum.
We find three kind of modifications. Firstly, the  energy
flux  fluctuates around its average value  with amplitudes
and frequencies determined by those of the metric fluctuations.
Secondly, this average value receives two positive contributions one
of which can be reinterpreted as  due to the `renormalisation' of the
surface gravity induced by the metric fluctuations. Finally, the
asymptotic spectrum is modified by the addition  of terms containing
thermal factors in which the frequency of the metric fluctuations
acts as a chemical potential.
\end{abstract}

\bigskip\bigskip

\baselineskip=.6cm

\newpage

\section{Introduction}
\setcounter{equation}0

25 years have passed since Hawking's theoretical 
discovery of quantum radiation
by black holes \cite{Hawk:75}.   Since then, many aspects of this
phenomenon have been investigated. First, the mean value of the
energy-momentum tensor has received much attention since it provides
the source of the so-called  semi-classical Einstein equations
(see e.g. \cite{BD,FrNo:98} and references therein). 
Hopefully, the solutions of these equations should  govern the mean
evolution of the evaporating geometry. More recently, more quantum
mechanical questions  which raise doubts concerning the validity of
this semi-classical evolution have also received attention.  In
particular, the controversial role  of arbitrarily large
(`transplanckian') frequencies of vacuum  fluctuations \cite{tHooft}--
\cite{CoJa:96}  
and the gravitational back reaction due to a
specific quantum \cite{KrWi, KKV} have been discussed. 

In this paper, we shall consider another aspect: We 
study how the fluctuations 
of the black hole 
horizon geometry might affect the properties of
Hawking radiation. To describe these fluctuations 
quantum mechanically and
to determine their effects on Hawking radiation
requires full quantum gravity. 
Besides the `spontaneous' metric fluctuations there also exist
so-called `induced' metric fluctuations, which are generated by quantum
fluctuations of all other fields interacting with the gravitational
one. In the regime when the `induced' metric fluctuations  are
dominating, a consistent way to describe black hole fluctuations and
back reaction is to use the stochastic semiclassical theory of gravity
based on the Schwinger-Keldysh effective action \cite{Schw:61,Keld:64}
and the Feynman-Vernon influence functional \cite{FeVe:63, FeHi:65}
methods. In stochastic gravity the semiclassical Einstein equations are
generalized to Einstein-Langevin equations which contain stochastic
stress-energy tensor describing metric fluctuations induced by
quantized fields. (For recent review see
\cite{Hu:99} -- \cite{MaVe:98b} and references therein.) 

The study  of the effects connected with black hole fluctuations is
a technically very complicated problem. Only some preliminary work has
been done in this direction till now. Under these conditions it is
natural to study simplified models. In particular,  it is not
unreasonable to hope that the main  properties of the Hawking radiation
modified by metric fluctuations can be extracted from a  much simpler
framework in which the fluctuations of the metric are treated
classically.

The model we shall use is inspired by that
proposed by York \cite{York:83}. In that 
model, the fluctuating geometry 
near the horizon of the black hole is represented by a 
Vaidya-type metric with a fluctuating mass.
The spectrum of these fluctuations is characterized by the
zero point fluctuations of quantum fields.
In this paper, we further simplify this model by considering only 
spherically symmetric fluctuations 
and by neglecting the 
scattering by the gravitational potential which occurs
in the 4-dimensional Dalembertian.
Then we determine how these fluctuations
modify the energy flux and the asymptotic spectrum
of s-waves.

The paper is organized as follows. York's model is described in
Section 2. Section 3 contains a perturbation analysis of the equation of
radial null ray propagation in the fluctuating geometry. 
The solution of this equation is obtained in Section 4
and used to obtain the modified energy flux  in Section 5 and 
the spectrum in Section 6.
The results are discussed in Section 7. In our work we use
dimensionless units where $G=c=\hbar=1$ and the sign conventions  of
\cite{MTW:73}.

\medskip

\section{Model}\label{s1.2}
\setcounter{equation}0

In order to study the  influence of metric fluctuations on  Hawking
radiation, we consider a simplified version 
of the model proposed  by York \cite{York:83}. In his model,  
the fluctuations of the black hole geometry are
approximated by an incoming Vaidya metric with a fluctuating mass. 
The fluctuating part of the mass function can be decomposed into
spherical harmonics.  Upon quantizing the gravitational field, only
components with $\ell\ge 2$ are important.  However quantum
fluctuations of matter fields  may induce fluctuations of the geometry
with all $\ell$.

In what follows we shall only consider spherical  modes of
fluctuations. Therefore, the metric for a neutral non-rotating black
hole  can be taken of the form
\be
ds^2 = -A\,dv^2+2dv\,dr+r^2dS_2^2\,, \n{2.1}
\label{metr}
\ee
where $dS_2^2$ is the metric of a unit 2-sphere and
\be
A = 1-\frac{2m}{r}\,, \n{2.2}
\ee
\be
m = m(v) = M[1+\mu (v)]\,\vartheta (v)\,, \n{2.3}
\ee

\be
\mu (v) = \mu_{0}\sin (\omega v)\,. \n{2.4}
\ee
This is the standard Vaidya metric in advanced time coordinates $(v,r)$. 
The function $\mu (v)$ encodes the fluctuations of dimensionless amplitude
$\mu _0$. The step function 
$\vartheta (v)$ in relation (\ref{2.3}) 
indicates that the black hole results from the gravitational collapse 
of a massive (with mass $M$) null shell propagating along $v=0$. 
Therefore inside the collapsing null shell the spacetime is flat. 

\begin{figure}
\centerline{\epsfig{file=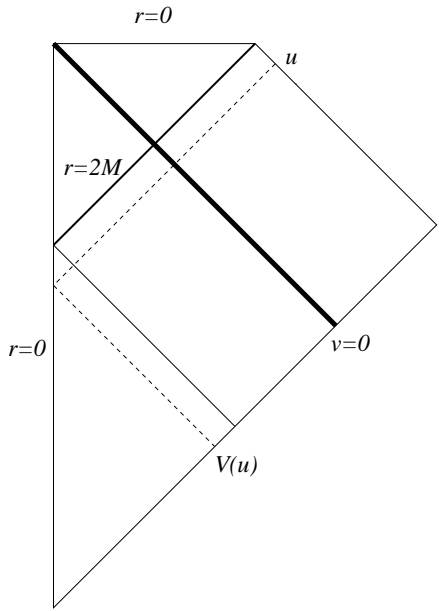, height=8cm}}
\caption[fig.1]{Conformal diagram for a black hole created by a
collapse
of a massive null shell. Solid dark line $v=0$ represents the
collapsing massive null shell.}
\label{fig.1}
\end{figure}

The conformal diagram of the whole geometry (\ref{2.1}) in the absence of
fluctuations (that is for $\mu_0=0$) is schematically shown in 
Figure \ref{fig.1}. The dashed line on this figure shows a
radial null ray which reaches ${\cal J}^+$ at the
moment of the retarded time $u$ and which was sent from ${\cal J}^-$
at $v=V(u)$ of advanced time. In the presence of
the fluctuations, the conformal structure of the spacetime remains
the same, but the function $V(u)$ is modified. We shall study this
modification in Section~4.

Upon substituting the metric (\ref{2.1}) into Einstein's equations, 
one easily gets that the right-hand side of these equations takes the form
\be
T_{\mu\nu} = \frac{1}{4\pi r^2}\,[M\delta (v) + M\mu (v)\,\vartheta (v)]
\,l_{\mu}\,{l}_{\nu}\,, \n{2.5}
\ee
where $l_{\mu} = -v_{,\mu}$ is a future-directed null vector tangent
to a radial in-going null ray. From a ``classical" viewpoint, 
$\mu (v)\,l_{\mu}\,l_{\nu}$ is a fluctuating flux of positive (for 
$2\pi n <\omega v <2\pi (n+1)$) and negative (for 
$2\pi (n+1) <\omega v <2\pi (n+2)$) energy density. Correspondingly,
the position of the apparent horizon $r_{\ind{AH}} = 2M(1+\mu (v))$  
also fluctuates near its average value $\bar{r}_{\ind{AH}} = 2M$.

Following York \cite{York:83} we assume that the dimensionless
amplitude $\mu_0=\alpha (m_{\ind{Planck}}/M)$, where $\alpha$ is a pure
number. 
In particular this assumption means that $\mu_0\ll 1$ for black
holes of mass $M\gg m_{\ind{Planck}}$.
It is also is assumed that in order to get a more realistic
result one should average over a spectrum of metric fluctuations.

\section{Radial Ray Propagation}\label{s3.1}
\setcounter{equation}0

\subsection{Radial null rays in perturbed geometry}
We first study the propagation of radial null rays in the 
fluctuating black hole geometry (\ref{2.1})
since we shall use, as usually done, geometric optics to 
construct the solutions of the wave equation.
 
In-going rays are given by $v = \mbox{const}$ 
and out-going rays obey the equation 
\be
A\,dv = 2\,dr\,. \n{3.1}
\ee
In order to solve this equation, 
we use a method of perturbations and write
\be
r = r(v) = R(v) + \rho (v)+ \sigma(v)+\ldots=
2M\left[ \tilde R(v) + \tilde \rho (v)+ \tilde 
\sigma(v) \right]+\ldots\, . \n{3.2}
\ee
$R(v)$ is the solution of  equation (\ref{3.1}) in the absence of
fluctuations, and $\rho (v)$ and $\sigma(v)$ 
are respectively the first  and second order perturbation 
in $\mu_0$. Higher order corrections are denoted by dots.
In what follows, we shall also often use the
dimensionless versions of $R$, $\rho$ and $\sigma$
which we mark by a tilde. 

The equation for out-going rays  in the unperturbed metric, 
[ $ (\ \ )^{\cdot } \equiv d/dv$ ]
\be
\dot{R}={1\over 2}\left(1 - \frac{2M\,\vartheta (v)}{R}\right)\, \n{3.3}
\ee
can be easily integrated.
Let us choose the value of the retarded time $u$ and denote by 
$r=R(v;u)$ the unperturbed  trajectory of a radial ray which arrives 
to ${\cal J}^+$ at the chosen time $u$.
This trajectory can be found by solving the equation
\be
u = v-2R_{\ast} = \mbox{const}\,. \n{3.4}
\ee
Here
\be
R_{\ast} = R-2M+2M\ln \frac{R-2M}{2M} \n{3.5}
\ee
is the usual tortoise radial coordinate.

The equations for the perturbations $\rho (v)$ and $\sigma(v)$
are obtained by linearizing (\ref{3.1}). 
Both functions obey the same equation
\be
\dot{f} = \frac{M}{R^2}\,f +F\,, \n{3.6}
\ee
for $v>0$ and 
\be
\dot{f} = 0\,, \n{3.7}
\ee
for $v<0$. For the first order perturbation, one has
\be
f=\rho\, , \hspace{1cm}F= - \frac{M}{R} \mu \, ,\n{3.8}
\ee
and for the second order perturbation
\be
f=\sigma\, ,\hspace{1cm}F=  \frac{M}{R^2} \mu \rho -\frac{M}{R^3} 
\rho^2\, .\n{3.9}
\ee
In these equations, 
 the retarded time $u$ is a fixed parameter which
specifies the unperturbed ray under consideration
and $R=R(v;u)$.

\subsection{Perturbed  horizon}

Before giving the general solution of the equations for the 
perturbations $\rho (v)$ and $\sigma(v)$ we
discuss the particular solution which describes the event horizon
in the fluctuating geometry.

First notice that $R=2M$ satisfies 
the unperturbed equation (\ref{3.3}). This degenerate solution
describes an outgoing null ray propagating along the unperturbed 
event horizon. Starting with this solution we easily obtain 
the following solutions for the dimensionless perturbations 
\be\n{3.9a}
\tilde \rho_{EH}= \mu_0\, \, {\Omega\,
\cos(\Omega\tilde{v})+\sin(\Omega\tilde{v})\over{1+\Omega^2}}\, ,
\ee
\be\n{3.9b}
\tilde \sigma_{EH}= \mu_0^2\, \, {2\Omega^2(2-
\Omega^2)\cos(2\Omega\tilde{v})
+\Omega(1-5\Omega^2)\sin(2\Omega\tilde{v})
\over{2(1+\Omega^2)^2(1+4\Omega^2)}}\, ,
\ee
where 
\be
\Omega=\omega/\kappa\, ,\hspace{0.5cm}
\tilde{v}=\kappa v\, , 
\ee
and $\kappa=(4M)^{-1}$ is the unperturbed surface gravity of the black
hole.

For both solutions it is possible to add a solution of the
homogeneous equation. Such a solution corresponds to radial null rays
propagating near the horizon in the unperturbed geometry. For these
solutions  the absolute value of $\rho$ and $\sigma$ is infinitely
growing. We choose  the integration constants  to exclude these
solutions so that the perturbed radial null ray neither goes to
infinity nor  to $r=0$. Therefore it describes the position of the
event horizon in the perturbed geometry. It is easy to verify that 
(\ref{3.9a}) and (\ref{3.9b})
coincides with the solution obtained by York \cite{York:83} [see
equations (4.8) and (4.9) of York's paper].

It is also interesting to  compute the modified value of 
the surface area ${\cal A}$ of the event horizon. When
averaging over time $v$, we find
\be\n{3.10a}
\bar{\cal A}\equiv 4\pi \overline{(r^2_{\ind{hor}}(v))}=
16\pi M^2\left[1+ {\mu_0^2\over{2(1+\Omega^2)}}\right]\, .
\ee
Similarly the average value of
the surface gravity in the fluctuating geometry is
\be\n{3.10b}
\bar{\kappa}\equiv \overline{\left({m(v)\over
r^{2}_{\ind{hor}}(v)}\right)}=\kappa\left[
1+3\overline{(\tilde{\rho}^2)}-2\mu_0\, 
\overline{(\tilde{\rho}\sin(\omega v))}
\right]
=\kappa \left[1+
{\mu_0^2\over{2(1+\Omega^2)}}\right]\, .
\ee
Upon computing the modifications of the Hawking
flux, we shall see that this `renormalized' surface
gravity will determined the modified temperature
$\bar{T}_H=\bar{\kappa}/2\pi$. The change of the area, $\delta {\cal A}=
\bar{\cal A}-{\cal A}$, and the change of temperature, $\delta
T_H=\bar{T}_H -T_H$, of the black hole induced by the metric
fluctuations obey the relation\footnote
{
It is interesting to note that  similar corrections to the black hole
surface area and temperature were obtained by Hu and Shiokawa
\cite{HuSh:97} in their stationary model of metric stochasticity. The
relation between $\delta {\cal A}$ and $\delta T_H$ in their case
contains an additional factor 2.
}
\be
{\delta {\cal A}\over {\cal A}}={\delta
T_H\over T_H}\, .
\ee

\subsection{Perturbed radial rays}

We now consider the general case, that is we assume that $R\ne 2M$.  
Equation (\ref{3.7}) can be easily integrated and
gives $f = \mbox{const}$  everywhere inside the collapsing shell. To
integrate equation (\ref{3.6}), we  change variable $v$ to $R(v;u)$,
where as earlier  the  retarded time parameter $u$ is fixed.
This allows us to rewrite  (\ref{3.6}) as
\be
\left(1-\frac{2M}{R}\right)\frac{df}{dR} - \frac{2M}{R^2}\,f = 
2F \,. \n{3.10}
\ee
The solution of this equation is 
\be
f = \left(1-\frac{2M}{R}\right)\left[-\int_R^{\infty}\frac{2F}{
\left(1-\displaystyle{\frac{2M}{R}}\right)^2}\,dR + f_0\right].
\n{3.11}
\ee
The integration constant $f_0$ corresponds to the  possibility to add
a solution of the homogeneous equation; that is, a solution of the 
equation (\ref{3.10}) with $F= 0$. We  put $f_0 = 0$
since this choice is equivalent to the requirement 
that the ray propagating in the
``perturbed"  geometry arrives to ${\cal J}^+$ at the same time $u$
as the ``unperturbed"  ray $r=R(v;u)$.

In what follows it is convenient to introduce 
the dimensionless quantities
\be
x = \frac{R-2M}{2M}\,, \quad \tilde{u} = \kappa u = \frac{u}{4M}\,, 
\n{3.12}
\ee
\be
x^{\ast} = \frac{R^{\ast}}{2M} = x+\ln x\,, \quad 
\tilde{f}=\frac{f}{2M}\,. 
 \n{3.13}
\ee 
In these notations solutions (\ref{3.11}) for the 
dimensionless perturbations 
$\tilde{\rho}$ and $\tilde{\sigma}$ take the form
\be
\tilde{\rho}(x)={x\over 1+x} I(x)\, ,\n{3.14}
\ee
\be
I(x)=\int_x^{\infty}{d\xi\over \xi^2}\, (1+\xi)\, \hat{\mu}(\xi)\, ,\n{3.15}
\ee
\be
\tilde{\sigma}(x)=-{x\over 1+x}\int_x^{\infty}{d\xi\over \xi^2}\,
\tilde{\rho}(\xi)\, \left(\hat{\mu}(\xi) 
-{\tilde{\rho}(\xi)\over 1+\xi}\right)\, .\n{3.16}
\ee
The fluctuating mass term $\hat{\mu} (\xi)\equiv
\mu(v(\xi))$ which enters these equations is
\be
\hat{\mu} (\xi)=\mu_0 \sin[\Omega(\xi+\ln\xi+\tilde{u})]\, .\n{3.17}
\ee

To study quantum black hole radiation in the fluctuating geometry, we
need to solve the wave equation in this geometry. As usual, we use
the geometrical optic approximation\footnote{
The validity of this approximation follows from 
the fact that the initial frequencies involved in the 
processes occuring at large $u$ times are much larger 
than the characteristic frequency $\kappa = 1/ 4M$.}. 
Thus we only need to solve the following problem \cite{Hawk:75}.
Consider a radial ray, which leaves $\cal{J}^-$ at some advance time
$v$ and reaches $\cal{J}^+$ at some retarded time $u$ (see Fig.~1). 
What is the relation between $u$ and $v$? To establish the relation
$v=V(u)$ we use the above  solution for a ray propagating in the
fluctuating geometry  outside the collapsing massive null shell, and
glue it to the solution inside the  shell. The latter means that the
values of $r$ coordinate for both rays must be the same on the shell
$v=0$. Using this condition  and the reflection at  $r=0$, one finds
for $\tilde{V}=V/4M$
\be
\tilde{V} = -[1+x_0+\tilde{\rho}(x_0)+\tilde{\sigma}(x_0)]\,, \n{3.18}
\ee
where  $x_0$  is the value of $x$ on the massive null shell (i.e. at
$v=0$). It depends on $u$  and  is a solution of the following equation 
\be
\tilde{u}+x_0+\ln x_0 = 0\, . \n{3.19}
\ee
We shall be interested in the null rays arriving to ${\cal J}^+$ at
late time $u$. For such rays the equation (\ref{3.19}) can be solved
by iterations and gives
\be
x_0 = e^{-\tilde{u}}\,(1- e^{-\tilde{u}}) + O( e^{-3\tilde{u}})\,.
\n{4.18}
\ee
This relation shows that for the late-time regime $x_0$ is very small.

\section{Calculation of $V(u)$-function}\label{s1.4}
\setcounter{equation}0

\subsection{First order corrections}

In this section we analyse the perturbations 
$\tilde \rho(x)$ and $\tilde \sigma(x)$ in the late-time regime.
Before proceeding to the computation of these perturbations 
it is appropriate to make a few remarks. 
First, we notice that it is necessary to 
compute $\tilde{\sigma}$, the quadratic fluctuation 
in $\mu_0$, since our aim is 
to obtain all quadratic corrections to the
flux and to the asymptotic spectrum.
Secondly, it should be noticed that the following 
developments for determining $V(u)$ in the 
perturbed geometry are quite similar to those of
refs. \cite{Bardeen, Massar} which concerned the 
determination of $V(u)$ when the energy-momentum
tensor of Hawking quanta is taken into account.

We start with the first order perturbation $\tilde \rho(x)$, see 
eq. (\ref{3.14}). To calculate $I(x)$ 
we notice that (\ref{3.15}) can be written in the form
\be
I(x) = \mu_0\,\mbox{Im}\left[PU_1(x;\Omega)+PU_2(x;\Omega)\right]\,, \n{4.1}
\ee
where for $n\ge 1$
\be\n{4.2a}
PU_n(x;\Omega)\equiv e^{i\Omega \tilde{u}}P_n(x;\Omega)
\ee
and
\be
P_n(x;\Omega) = \int_x^{\infty}
\frac{e^{i\Omega \xi }}{\xi^{n-i\Omega}}\,d\xi\,. \n{4.2}
\ee
By integrating by parts, it is easy to show that
\be
P_n(x;\Omega) = \frac{e^{i\Omega x }}{(n-1-i\Omega)\,x^{n-1-i\Omega}} 
+ \frac{i\Omega}{n-1-i\Omega}\,P_{n-1}(x;\Omega)\,. \n{4.3}
\ee
Using this relation, we can rewrite (\ref{4.1}) as
\[
I(x) =\mu_0\,\mbox{Im}\left\{\frac{1}{1-i\Omega}
\left[e^{i\Omega (\tilde{u}+x+\ln x)}{1\over {x}}
 + \,PU_1(x;\Omega)\right]\right\}
\]
\be
\hspace{.9cm} = \frac{\mu_0}{x\sqrt{1+\Omega^2}}\,\sin [\Omega (\tilde{u}
+x+\ln x) + \arctan \Omega ] + \tilde{I}(x)\,, \n{4.4}
\ee
where
\be
\tilde{I}(x) = \mu_0\,\mbox{Im}\left[\frac{PU_1(x;\Omega)}{1-
i\Omega}\right]\,. \n{4.5}
\ee
Function $P_1(x;\Omega)$ can be expressed in terms of the incomplete 
gamma 
function $\Gamma (\alpha ,\zeta )$. Using the relation 
\be
\int_u^{\infty}\frac{e^{-\mu \xi }}{\xi^{1-\alpha}}\,d\xi = \mu^{-\alpha}\,
\Gamma (\alpha ,\mu u)\,, \n{4.6}
\ee 
we get
\be
P_1(x;\Omega) = (-i\Omega )^{-i\Omega}\,\Gamma (i\Omega ,-i\Omega x) = 
e^{-\frac{\pi\Omega}{2} - i\Omega \ln \Omega }\,\Gamma (i\Omega ,
-i\Omega x)\,. \n{4.7}
\ee
Furthermore, the incomplete gamma function $\Gamma (\alpha ,\zeta )$ 
allows the 
following series expansion
\be
\Gamma (\alpha ,\zeta ) = \Gamma (\alpha )\left[1 - \zeta ^{\alpha}\,
e^{-\zeta}\,\sum_{n=0}^{\infty}\frac{\zeta^n}{\Gamma (\alpha 
+n+1)}\right]. 
\n{4.8}
\ee
Thus for small $x$ ($\Omega x\ll 1$) we have
\be
\Gamma (i\Omega ,-i\Omega x) \approx  \Gamma (i\Omega ) + \frac{i}
{\Omega}\,e^{\frac{\pi\Omega}{2}}\,x^{ i\Omega }\,e^{i\Omega \ln \Omega 
}\,, 
\n{4.9}
\ee
and
\be
P_1(x;\Omega) \approx e^{-\frac{\pi\Omega}{2} - i\Omega \ln \Omega 
}\,\Gamma 
(i\Omega ) + \frac{ix^{i\Omega}}{\Omega}\,. \n{4.10}
\ee

For the calculation of $V(u)$ we need to know $I(x_0)$ where $x_0$ is 
the solution of equation (\ref{3.19}).
Using (\ref{4.4}), one gets 
\be
I(x_0) = \frac{\mu_0\,\Omega }{x_0(1+\Omega^2)} + \tilde{I}(x_0) \,, \n{4.11}
\ee
where
\be
\tilde{I}(x_0) = \mu_0\,\mbox{Im}\left[\frac{ie^{-i\Omega x_0}}{\Omega
(1-i\Omega )}\right]  +  
\mu_0\,\mbox{Im}\left[e^{-\frac{\pi\Omega}{2}-i\Omega 
\ln \Omega+i\Omega\tilde{u}}\,{\Gamma (i\Omega )\over
{1-i\Omega}}\right]\, . \n{4.12}
\ee
For small $x_0$ the first term in the right-hand side gives $\mu_0 
/[\Omega 
(1+\Omega^2)]$. To calculate the second term contribution, we notice that 
\be
e^{-\frac{\pi\Omega}{2}-i\Omega 
\ln \Omega}\Gamma (i\Omega ) = 
q(\Omega)\,
e^{-i\varphi_{\Gamma}(\Omega )}\,, \n{4.13}
\ee
where 
\be\n{4.14}
q(\Omega)=\frac{\sqrt{2\pi}}{\sqrt{\Omega (e^{2\pi\Omega}-1)}}\, ,
\ee
and $\varphi_{\Gamma}(\Omega )$ is a real function which for large values 
of $\Omega ,\ \Omega \rightarrow \infty $, has the following expansion
\be
\varphi_{\Gamma}(\Omega ) \approx  \Omega +\frac{1}{4}
\,\pi + 
\sum_{n=1}^{\infty}\frac{(-1)^{n-1}\,B_{2n}}{(2n-1)(2n)\Omega^{2n-1}}\,, 
\n{4.14a}
\ee
where $B_n$ are Bernoulli numbers.

Inserting (\ref{4.13}) into (\ref{4.12}) one gets for small $x_0$
\be
\tilde{I}(x_0) \approx \frac{\mu_0}{\Omega (1+\Omega^2)} + \mu_0\,
\frac{q(\Omega)}{\sqrt{1+\Omega^2}}\, \sin \Phi_1\,, \n{4.15}
\ee
where 
\be
\Phi_1 = \Phi(\Omega)  +\arctan \Omega\, ,
\hspace{0.5cm}\Phi(\Omega)=\Omega\tilde{u} - 
\varphi_{\Gamma }(\Omega )\, .
\n{4.16}
\ee

Collecting all the results, we finally obtain
\be
I(x_0) \approx \mu_0\left[\,\frac{\Omega }{x_0(1+\Omega^2)} + \frac{1}
{\Omega (1+\Omega^2)}
+ \,\frac{q(\Omega)}{\sqrt{1+\Omega^2}}\,\sin \Phi_1\,\right]\, . \n{4.17}
\ee
This result can be used to obtain the asymptotic form of the
function  $\tilde{V} (u)$.  Using equations (\ref{3.14}),
(\ref{3.18}),  (\ref{4.17}), and (\ref{4.19}), we find up  to the
first order in $ e^{-\tilde{u}}$
\be
-\tilde{V} \approx \hat{V}_0 + e^{-\tilde{u}}\,[\hat{C}_1 + \hat{C}_2\sin
(\Phi_1)] 
+ \tilde{V}_2 (\tilde{u})\,, \n{4.19}
\ee
where
\be
\hat{V}_0 = 1 + \mu_0\, \frac{\Omega }{1+\Omega^2}\,, 
\hspace{0.5cm}
\hat{C}_1 = 1 - \mu_0\,\frac{\Omega ^2-1}{\Omega (1+\Omega^2)} \,, 
\hspace{0.5cm}
\hat{C}_2 =  \mu_0\,\frac{q(\Omega)}{\sqrt{1+\Omega^2}} \,, 
\n{4.22}
\ee
and
\be
\tilde{V}_2(\tilde{u}) = \tilde{\sigma}(x_0(\tilde{u})) \n{4.23}
\ee
is the second order correction.

\subsection{Second order corrections}\label{s4.1.1} 

To calculate the second order correction $\tilde{\sigma}$, we must 
estimate the integrals of eq. (\ref{3.16}).
By integrating by parts, $\tilde{\sigma}$ can be rewritten as
\be
\tilde{\sigma}(x)={x\over 1+x}\left( {I^2(x)\over 2(1+x)^2}
-\int_x^{\infty}{d\xi\over \xi^2} \hat{\mu} (\xi) I(\xi)\right) \,
,\n{4.24}
\ee
where as earlier
\be
I(x) = I_1(x) + I_2(x)\, , \n{4.25}
\ee
and
\be
I_n(x) = \int_x^{\infty}\frac{d\xi}{\xi^n}\,\hat{\mu}(\xi )\,. \n{4.26}
\ee
It is easy to see that
\be
\int_x^{\infty}\frac{d\xi}{\xi^2}\,\hat{\mu}(\xi )\,I_2(\xi) 
= \frac{1}{2}[I_2(x)]^2 \,. 
\n{4.27}
\ee
Thus we have
\be
\tilde{\sigma}(x) = \frac{x}{1+x}\left[\frac{(I_1(x)+I_2(x))^2}{2(1+x)^2} 
- 
\frac{1}{2}[I_2(x)]^2 - T(x)\right],  \n{4.28}
\ee
where
\be
T(x) = \mu_0\,\mbox{Im}\left[e^{i\Omega\tilde{u}}\,Q_2(x;\Omega
)\right] \, ,
 \n{4.29}
\ee
and
\be
Q_n(x;\Omega ) = \int_x^{\infty}\frac{d\xi}{\xi^{n-i\Omega}}
I_1(\xi) \,. \n{4.30}
\ee
The functions $I_n(x)$ are related to $PU_n(x;\Omega )$ defined by
equation  (\ref{4.2a}) as follows
\be
I_n(x) = \mu_0\,\mbox{Im}\left[PU_n(x;\Omega )\right]. 
 \n{4.31}
\ee
Using the results of the previous section, we can obtain the
expansions of  $I_n(x_0)$ for small $x_0$. Thus, the problem of
finding $\tilde{\sigma}(x_0)$  for small $x_0$ is reduced to study
the function $T(x)$.  For this purpose we first obtain a recursion
relation for $Q_n(x;\Omega )$.  By integrating by parts, we have
\[
Q_n(x;\Omega ) = \frac{e^{i\Omega x}\,I_1(x)}{(n-1-i\Omega )
\,x^{n-1-i\Omega}}
- \frac{1}{(n-1-i\Omega )}\,\int_x^{\infty}
\frac{d\xi\,e^{i\Omega\xi}\,\hat{\mu}
(\xi )}{\xi^{n-i\Omega}}
\]
\be
\hspace{2.5cm} +\,\frac{i\Omega}{n-1-i\Omega }\,Q_{n-1}(x;\Omega )  \,. 
\n{4.32}
\ee
Using (\ref{3.17}) and the definition (\ref{4.2a}) of $PU_n(x;\Omega)$
we can write the second term of (\ref{4.32}) as
\be
\int_x^{\infty}\frac{d\xi\,e^{i\Omega\xi}\,
\hat{\mu}(\xi )}{\xi^{n-i\Omega}} = 
-{i\over 2}\mu_0 \, e^{-i\Omega\tilde{u}}
\left[PU_n(x;2\Omega ) - \frac{1}{(n-1)\,x^{n-1}}\right]. \n{4.33}
\ee
Relations (\ref{4.32}) and (\ref{4.33}) allow us to write  
 $Q_2(x;\Omega ) $ as
\[
Q_2(x;\Omega ) = \frac{e^{i\Omega x}\,I_1(x)}{(1-i\Omega )\,x^{1-i\Omega}}
-\frac{\mu_0 \,e^{-i\Omega \tilde{u}}}{2i(1-i\Omega )}
\left[PU_2(x;2\Omega ) - 
\frac{1}{x}\right] 
\]
\be
\hspace{2,34cm} +\,\frac{i\Omega}{1-i\Omega }\,Q_{1}(x;\Omega )  \,. 
\n{4.34}
\ee
We also have
\[
Q_1(x;\Omega ) = \frac{\mu_0}{2i}
\left[\,e^{i\Omega\tilde{u}}\,\int_x^{\infty}
\frac{d\xi\,e^{i\Omega\xi}}{\xi^{1-i\Omega}} \int_{\xi}^{\infty}
\frac{d\eta\,e^{i\Omega\eta}}{\eta^{1-i\Omega}}
 -\,e^{-i\Omega\tilde{u}}\,\int_x^{\infty}
\frac{d\xi\,e^{i\Omega\xi}}{\xi^{1-i\Omega}} \int_{\xi}^{\infty}
\frac{d\eta\,e^{-i\Omega\eta}}{\eta^{1+ i\Omega}}\,\right]
\]
\be
\hspace{1.6cm} =  \frac{\mu_0}{2i}\, e^{-i\Omega\tilde{u}}\,
\left[\,\frac{1}{2}\,\left(PU_1(x;\Omega )\right)^2 - S(x;\Omega )\,
\right] , \n{4.35}
\ee
where
\be
S(x;\Omega ) = \int_x^{\infty}
\frac{d\xi\,e^{i\Omega\xi}}{\xi^{1-i\Omega}} 
\int_{\xi}^{\infty}
\frac{d\eta\,e^{-i\Omega\eta}}{\eta^{1+ i\Omega}}\,. \n{4.36}
\ee
This function allows the following representation (see Appendix A)
\[
S(x;\Omega ) = \frac{i}{\Omega}\,e^{- \frac{\pi\Omega}{2} + i\Omega\ln 
\Omega }\,\Gamma(-i\Omega )\,x^{i\Omega}\,{}_1F_1(i\Omega ;1+ i\Omega ;
i\Omega x)
\]
\be
\hspace{1.5cm} +\, \frac{i}{\Omega}\left[\ln (i\Omega x) + f(-i\Omega ;
i\Omega x) - \psi(i\Omega )\right],  \n{4.37}
\ee
where ${}_1F_1$ is a hypergeometric function, 
$\psi (z) = (d\ln \Gamma (z)/dz)$, and
\be
f(\alpha ;y) = \sum_{n=1}^{\infty}\frac{y^n}{n(1+\alpha )_n}\,. \n{4.38}
\ee
Here
\be
(1+\alpha )_n = \frac{\Gamma (1+\alpha +n)}{\Gamma (1+\alpha )}\,. \n{4.39}
\ee
Combining these results, we can rewrite (\ref{4.34}) as
\[
e^{i\Omega\tilde{u}}\,Q_2(x;\Omega ) = \frac{e^{i\Omega (\tilde{u}+x +\ln 
x)}\,
I_1(x)}{(1-i\Omega )\,x} 
 -\,\frac{\mu_0}{2i(1-i\Omega )}\left[
PU_2(x;2\Omega ) - \frac{1}{x}\right]
\]
\be
\hspace{2.4cm} +\,\frac{\mu_0\,\Omega}{2(1-i\Omega )} \left[\frac{1}{2}\,
\left( PU_1(x;\Omega )\right)^2 - S(x)\right] . \n{4.39a}
\ee
The final expression for the second order perturbation
$\tilde{\sigma}(x_0)$ is obtained by inserting (\ref{4.29}),
(\ref{4.39a}), and (\ref{4.31}) into (\ref{4.28}).

Since we are  interested in $\tilde{\sigma}(x_0)$ where $x_0$ is
small we need only to know the first terms of its expansion in powers
of $x_0$. To do this we use  the
following asymptotics at small value of $x_0$
\be
PU_1(x_0;\Omega)\approx {i\over \Omega}+q(\Omega)e^{i\Phi(\Omega)}\, ,
\ee
\be
PU_2(x_0;\Omega)\approx {1\over 1-i\Omega}\left[ {1\over x_0}+i\Omega
PU_1(x_0;\Omega)\right]\, ,
\ee
\be
S(x_0;\Omega)\approx {i\over\Omega}\left[ q(\Omega) e^{-i\Phi(\Omega)}
+\ln(\Omega x_0) +{i\pi\over 2}-\psi(i\Omega)\right]\, .
\ee
The approximative values of $I_1(x_0)$ and $I_2(x_0)$ can be easily
obtained by using the relation (\ref{4.31}).
The  calculations the asymptotic value of $\sigma(x_0)$ is hence
straightforward but quite long. We use Maple  to perform these
calculations. The result is
\be \n{4.40}
\tilde{\sigma}(x_0)\approx \mu_0^2\left[\sigma_0 +\sigma_1 x_0 
+\sigma'\right]\, ,
\ee
where
\be \n{4.41}
\sigma_0={\Omega^2(2-\Omega^2)\over {(1+\Omega^2)^2(1+4\Omega^2)}}\, ,
\ee
\be \n{4.42}
\sigma_1=\sigma_1^0+\sigma_1^1 \, q(\Omega)+\sigma_1^2 \,
q^2 (\Omega)+\sigma_1^3 \, q(2\Omega)\, 
\ee
\be \n{4.43}
\sigma_1^0= {2+3\Omega^2-33\Omega^4+20\Omega^6
\over{4\Omega^2(1+\Omega^2)^2(1+4\Omega^2)}}\,
\ee
\be \n{4.44}
\sigma_1^1= {(\Omega^2-1)[\Omega
\cos(\Phi(\Omega))+\sin(\Phi(\Omega))]\over {\Omega(1+\Omega^2)^2}}\,
,
\ee
\be \n{4.45}
\sigma_1^2={(1-\Omega^2)+[\Omega\sin(2\Phi(\Omega))-
\cos(2\Phi(\Omega))]\over{4(1+\Omega^2)}}\, ,
\ee
\be \n{4.46}
\sigma_1^3=-{\Omega\left[(2\Omega^2-1)\sin(\Phi(2\Omega))
-3\Omega\cos(\Phi(2\Omega))\right]\over{(1+\Omega^2)(1+4\Omega^2)}}
\, ,
\ee
and
\be
\sigma'=-{x_0\over 4(1+\Omega^2)}\left[ \pi\Omega -2\ln(\Omega x_0)-2
\mbox{Im} [(\Omega -i)\psi(i\Omega)]\right]\, .
\ee

Let us recall that in these expressions $x_0$ is a function of the
retarded time $u$ given by (\ref{4.18}). We can now include the second
order corrections (\ref{4.23}) into the expression  for
$\tilde{V}(u)$ and write it in a form similar to
(\ref{4.19})
\be
-\tilde{V} \approx \tilde{V}_0 + e^{-\tilde{u}}\,\left[C_1 + C_2\sin
(\Phi_1) +C_3\sin(\Phi_2)+C_4\sin(\Phi_3)+C\tilde{u}\right] 
\,, \n{4.48}
\ee
where
\be\n{4.49}
\tilde{V}_0 = 1 + \mu_0\, \frac{\Omega }{1+\Omega^2}
+\mu_0^2 {\Omega^2(2-\Omega^2)\over {(1+\Omega^2)^2(1+4\Omega^2)}}
\,, 
\ee
\[
C_1 = 1 - \mu_0\,\frac{\Omega ^2-1}{\Omega (1+\Omega^2)} 
+\mu_0^2 \left[ {2+3\Omega^2-33\Omega^4+20\Omega^6
\over{4\Omega^2(1+\Omega^2)^2(1+4\Omega^2)}}\right.
\]
\be\n{4.50}
\left.
+{1-\Omega^2
\over{4(1+\Omega^2)}}q^2(\Omega)-{(\pi\Omega-2\ln\Omega
-2\mbox{Im}[(\Omega-i)\psi(i\Omega)])\over{4(1+\Omega^2)}}
\right]
\,, 
\ee
\be\n{4.51}
C_2 =  \mu_0\,\frac{q(\Omega)}{\sqrt{1+\Omega^2}}
\left[ 1+\mu_0{\Omega^2-1\over{\Omega(1+\Omega^2)}}\right] \,, 
\ee
\be\n{4.52}
C_3 =  \mu_0^2{q^2(\Omega)\over{4\sqrt{1+\Omega^2}}}\, ,
\ee
\be\n{4.53}
C_4 = - \mu_0^2{\Omega\, 
q(2\Omega)\over{\sqrt{(1+\Omega^2)(1+4\Omega^2)}}}\, ,
\ee
\be\n{4.54}
C = - {\mu_0^2\over 2(1+\Omega^2)}\, ,
\ee
and the phases $\Phi_i$ are defined as follows
\be\n{4.55}
\Phi_1=\Omega \tilde{u}-\varphi_{\Gamma}(\Omega)+\arctan\Omega\, ,
\ee
\be\n{4.56}
\Phi_2=2\Omega \tilde{u}-2\varphi_{\Gamma}
(\Omega)-\arctan({1\over\Omega})\, ,
\ee
\be\n{4.57}
\Phi_3=2\Omega
\tilde{u}-\varphi_{\Gamma}(2\Omega)-
\arctan({3\Omega\over2\Omega^2-1})\, .
\ee

Notice, that the coefficients $\hat{V}_0$, $\hat{C}_1$ and $\hat{C}_2$
which appeared in the first order expression (\ref{4.19}) 
are now replaced in (\ref{4.48}) by 
the new coefficients $\tilde{V}_0$, $C_1$ and $C_2$  
with the only difference that the corresponding
coefficients get second order corrections. Notice also that in 
(\ref{4.48}),  
the terms with double frequency $2\Omega$ 
and a term which is linear in $\tilde{u}$ are new
with respect to the first oder result (\ref{4.19}).

\section{Calculation of the Flux of Hawking Radiation}
\setcounter{equation}0

Now we derive the $s$-mode contribution to Hawking radiation. In what
follows, we shall neglect the scattering  by the gravitational
potential barrier which appears in the 4D Dalembertian.  In other
words, we use 2D approximation
in which ingoing and outgoing modes completely decouple.
This strong hypothesis requires some explanations.
The decoupling of the modes greatly simplifies the 
calculation of the asymptotic flux when the metric
is no longer static. Indeed, the height of the potential barrier
now depends on $v$ in the metric eq. (\ref{metr}).
Therefore one looses the fact that the transmission 
coefficients are diagonal in energy. Moreover,
the new coefficients will also mix positive and negative
frequency modes. This will lead to additional 
pair creation probabilities. Thus there will be interference
effects between the usual pair creation amplitudes induced by the 
frequency mixing governed by eq. (\ref{4.48})
and these new coefficients.

To determine the importance of these interesting effects
is complicated and goes beyond the scope of the present paper
which is to describe the effects on Hawking radiation 
induced by the fluctuations of the geometry in the very 
close vicinity of the horizon. In this respect,
we wish to emphasize that our classical metric has been
choosen to mimic the near horizon quantum fluctuations
and not the fluctuations of the height of the barrier around
$r= 3M$. On physical grounds, in a self-consistent treatment,
one might expect that the residual fluctuations around $3M$
be much smaller that the near horizon ones. Therefore the neglection
of the time dependence of the barrier might turn out to 
be physically legitimate.

In the 2 dimensional simplified description,
when the field is in its vacuum
state before the formation of the black hole, 
the mean energy flux  at ${\cal J}^+$ is
\be
{dE\over du}\equiv 4\pi r^2 \langle T_{uu}\rangle^{\ind{ren}}
=
\frac{\kappa^2}{12\pi}\left(
\frac{d\tilde{V}}{d\tilde{u}}\right)^{1/2}\,
\frac{d^2\
}{d\tilde{u}^2}
\left[\left(\frac{d\tilde{V}}{d\tilde{u}}\right)^{-1/2}\right].
 \n{5.1}
\ee
Here $\tilde{V}(\tilde{u})$ is the function calculated in the 
previous section. Notice also, that $u$ in this relation  is the
proper time at ${\cal J}^+$ in the perturbed geometry,
see the remark made after eq. (\ref{3.11}).
Thus this is the time which defines positive frequency at ${\cal J}^+$.

Before presenting the results of the calculations of $dE/du$ we make
several remarks.  First, it is evident that the expression for
$dE/du$ does not depend on the value  of constant $\tilde{V}_0$ in
equation (\ref{4.48}). For this reason we can put it equal to zero.
This corresponds to a simple redefinition $\tilde{V}\rightarrow
\tilde{V}+\tilde{V}_0$. Moreover, $dE/du$ is not changed if we
multiply $\tilde{V}$ by an arbitrary constant.  For these reasons 
the calculations of $dE/du$ can be  performed  with the simplified
form for $\tilde{V}$ 
\be
\tilde{V} = -e^{-\tilde{u}}\,[1 + A_1\sin (\Omega \tilde{u} +
\varphi_1 )+A_2\sin (2\Omega \tilde{u} +
\varphi_2 )+C\tilde{u}]\,. \n{5.3}
\ee
In this expression, we have kept all terms up to second order in
$\mu_0$ and introduced the following notations
\be\n{5.4}
A_1\equiv
\mu_0 a_1+\mu_0^2 b_1=\mu_0\,\frac{q(\Omega)}{\sqrt{1+\Omega^2}}
\left[ 1+2\mu_0{\Omega^2-1\over{\Omega(1+\Omega^2)}}\right] \,, 
\ee
\be\n{5.5}
\varphi_1=-\varphi_{\Gamma}+\arctan \Omega\, .
\ee
In the same way the two terms in (\ref{4.48}) with coefficients $C_3$
and $C_4$ having the same dependence $2\Omega\tilde{u}$ on the retarded
time $\tilde{u}$ have been combined into the following single term
\be
A_2\sin (2\Omega \tilde{u} +\varphi_2 )\equiv
C_3\sin(\Phi_2)+C_4\sin(\Phi_3)
\label{5.3d}
\ee
$A_2$ is of second order in $\mu_0$ and, with notations to 
(\ref{5.4}), can be written as $A_2\equiv \mu_0^2\, b_2$.
The explicit expressions for $b_2$ and $\varphi_2$
can be obtained easily, but since they are very long and are not
important for our final result we do not reproduce them.
Note also, that
since $C$ is already a quantity of  second order in $\mu_0^2$, the
redefinition of $\tilde{V}$  does not affect it.

The calculation of $dE/du$ from (\ref{5.1}) and (\ref{5.3}) is
straightforward but long and was performed by using Maple. 
It is convenient to write the result in the form
\be\n{5.6}
{dE/du}= (dE/du)^{\ind{perm}}+(dE/du)^{\ind{fluct}}\, ,
\ee
where $(dE/du)^{\ind{perm}}$ is the  mean value of the flux and
$(dE/du)^{\ind{fluct}}$ is its fluctuating part. The latter will of
course not contribute to the total energy received on ${\cal J}^+$.

The constant part is 
\be\n{5.7}
(dE/ du)^{\ind{perm}}=
{\kappa^2\over 48\pi}\left[1+{1\over 2}\mu_0^2\Omega^2\,
q^2(\Omega)-2C\right]\, ,
\ee
and the fluctuating part is
\[
(dE/ du)^{\ind{fluct}}=
-{\mu_0\over 2}\Omega\sqrt{1+\Omega^2}q(\Omega)\cos(\Omega
\tilde{u}+\varphi_1)
\]
\[
+\mu_0^2\left[
q(\Omega){1-\Omega^2\over{\sqrt{1+\Omega^2}}}\, 
\cos(\Omega\tilde{u}+\varphi_1)
+q^2(\Omega){\Omega^2(1-5\Omega^2)\over{8(1+\Omega^2)}}\,
\cos(2\Omega\tilde{u}+2\varphi_1)
\right.
\]
\be\n{5.8}
\left.
+q^2(\Omega){\Omega(1+4\Omega^2)\over {4(1+\Omega^2)}}\,
\sin(2\Omega\tilde{u}+2\varphi_1)
-b_2 \Omega(1+4\Omega^2)\, \cos(2\Omega\tilde{u}+\varphi_2)
\right]\, .
\ee
The function $q(\Omega)$ is given by (\ref{4.14}) and, in the last
term,  $b_2 \cos(2\Omega\tilde{u}+\varphi_2)$  is equal to
\be
b_2\cos(2\Omega\tilde{u}+\varphi_2)=
{q^2(\Omega)\over{4\sqrt{1+\Omega^2}}}\cos(\Phi_2) - {\Omega\,
 q(2\Omega)\over{\sqrt{(1+\Omega^2)(1+4\Omega^2)}}}\cos(\Phi_3)\, .
\ee

The remarkable fact  is that, to second order in $\mu_0$, the
correction term which is linear in $\tilde{u}$ in (\ref{5.3}) does not 
give any
time-dependent contribution. It only gives an additional constant to
${(dE/du)}^{\ind{perm}}$ in (\ref{5.7}).  
This has the following simple explanation:
the term $C\tilde{u}$ can be removed from (\ref{5.3}) by absorbing it
into $e^{-\tilde{u}}$ without  changing the other terms in our second
order expression. This transformation corresponds to the
``renormalization'' of the surface gravity
\be\n{5.10}
\kappa \rightarrow
\kappa_{r}=\kappa(1-C)=\kappa\left[1+{\mu_0^2\over{2(1+\Omega^2)}}
\right]\, .
\ee
Hence the expression for $(dE/du)^{\ind{perm}}$ can be identically
rewritten as
\be\n{5.11}
(dE/du)^{\ind{perm}}={\kappa_{{r}}^2\over 48\pi}\left[1+{1\over
2}\mu_0^2\Omega^2\, q^2(\Omega)\right]=
{\kappa_{{r}}^2\over
48\pi}\left[1+\mu_0^2{\pi\Omega\over{\exp(2\pi\Omega)-1}}\right]
\, .
\ee
It is interesting to note that the renormalized surface gravity
$\kappa_{{r}}$ which is introduced here  coincides with the
average value of the surface gravity $\bar{\kappa}$ defined by
equation (\ref{3.10b}).

\section{The Modified Asymptotic Spectrum}
\setcounter{equation}0

Instead of focusing on integrated quantities, we shall now consider
how the asymptotic spectrum is modified by  the fluctuating part of
the metric. As usual, the asymptotic  spectrum is characterized by
the angular momentum (that we take to zero) and the asymptotic
energy, $\lambda$, the eigenvalue of $i\partial_u$. Indeed the 
fluctuations we are considering do not affect the  stationary
character of the asymptotic (large $r$) metric.

To obtain the modified spectrum,  we need to compute the Bogoliubov
coefficients in the modified geometry. Before proceeding to this
calculation, it should be noticed that the 2D expressions we shall
use are exact only for large $\la$, i.e. $\kappa \la \gg 1$. At lower
frequencies, there is indeed a potential barrier in the  4D Dalembertian
for s-waves which reduces the transmitted flux
in a static geometry, c.f. the discussion at the beginning
of the former Section.

We first recall how the well known properties of  Hawking spectrum are
extracted from  the Bogoliubov coefficients. The latter are given by
the overlap of the initial (infalling) modes which are specified on
${\cal J}^-$ and the final (outgoing) modes   specified on ${\cal
J}^+$. Both are solutions of the Dalembertian  equation in the metric
(\ref{2.1}). For s-waves and under the neglection of the potential
barrier, these modes satisfy the  2D equation $\partial_u \partial_v
\phi = 0$. Thus the in-modes can be decomposed in terms of planes
waves 
\be
\phi_\nu(v) = {e^{-i\nu v} \over \sqrt{4 \pi \nu} }
\label{inm}
\ee
where $\nu$ is the energy measured on ${\cal J}^-$.
Similarly, the out-modes are
\be
\phi_\lambda(u) = {e^{-i\lambda u} \over \sqrt{4 \pi \lambda} }
\label{outm}
\ee
where $\la$ is the energy measured on ${\cal J}^+$.

The scattering of in-modes in the time dependent geometry simply
follows from the `reflection' condition on $r=0$ wherein the
Wronskian must vanish. This implies that the scattered in-modes are
given by $\phi_\nu(V(u))$.
Then the Bogoliubov coefficients are 
\ba
\al_{\nu, \lambda} &=& \int \! du
\; \phi^*_\nu(V(u))\;
i\! \lr{\partial_u}\;
\phi_\lambda(u) 
=
\int\! du {e^{i\nu V(u)} \over \sqrt{4 \pi \nu} }
{e^{-i\lambda u} \over \sqrt{ \pi \lambda^{-1}} }
\nonumber\\
\beta_{\nu, \lambda}&=&  \int \! du \;
\phi_\nu(V(u))\;
i\! \lr{\partial_u} \;
\phi_\lambda(u) 
= \int\! du {e^{-i\nu V(u)} \over \sqrt{4 \pi
\nu} } 
{e^{-i\lambda u} \over \sqrt{ \pi \lambda^{-1}} }
\label{abcoef}
\ea
The second expressions follow  from 
integration by part and the neglect of the 
end-point contributions.

In the unperturbed geometry, for large $u$, $\kappa V(u)$ is given
by  $-1- e^{-\kappa u}$, see eq. (\ref{3.18}). The validity of this
asymptotic expression  requires that the initial frequencies $\nu$ be
large enough. Indeed, upon applying  the stationary phase condition
to the integrand of the coefficient $\al_{\nu, \lambda}$ one obtains
the following relation for the position of the saddle point $u_{sp}$
\be
 \lambda = \nu e^{-\kappa u_{sp}}\, . 
\label{resc}
\ee
This equation simply specifies the value of $u$ at which the Doppler
shifted initial frequency $\nu$  resonates with $\lambda$.  Notice
in passing that  it is this exponentially small Doppler factor which
leads to the necessity of considering arbitrary large initial
(`transplanckian') frequencies. From  eq. (\ref{resc}), one deduces 
that large $\nu$ means that the corresponding value of $e^{-\kappa
u_{sp}}$ satisfies  $e^{-\kappa u_{sp}} \gg e^{-2\kappa u_{sp}}$.
Notice also that the  stationary phase condition applied to the
$\beta$ coefficient,  leads to the same condition  up to an overall
negative sign.  This implies that the location of the saddle 
point in the complex $u$-plane receives
an  imaginary contribution equal to $-i \pi /\kappa$. The
determination of the sign of this imaginary part follows from the
fact that $V(u)$ appearing in $e^{-i \nu V}$  must belong to the
lower half complex $V$-plane. Physically this amounts to specify that the
in-vacuum   contains no excitation characterized by positive $\nu$.
Mathematically it gives $\vert \beta_{\nu, \lambda} / \alpha_{\nu,
\lambda} \vert^2 = e^ {- 2 \pi \la /\kappa}$.

These considerations based on a saddle point analysis are confirmed
by the exact integration of eqs. (\ref{abcoef}). It will be found
useful to express the exact expressions in terms of the following
function
\be
B(\nu, \lambda) = 
\Gamma( i{ \la \over \kappa}) \ \sqrt{ \la \over ( 2\pi \kappa  )}
({\nu \over \kappa })^{- i \la /\kappa} 
\ e^{- \pi \la /2 \kappa }\, ,
\label{Bd}
\ee
where $\Gamma(x)$ is the Euler complete gamma 
function. The norm of this function
gives: $\vert B\vert ^2 =(e^{2 \pi \la / \kappa } - 1)^{-1}$. 
Upon extending the domain of 
validity of the asymptotic behaviour of $V(u)$
for all $u$, one finds
\ba
\al_{\nu, \lambda} &=& B(\nu, \lambda) 
\sqrt{ 1 \over 2 \pi  \kappa \nu} 
\ e^{\pi \la /\kappa } e^{-i \nu/\kappa}\, ,   
\nonumber\\
\beta_{\nu, \lambda} &=& 
B(\nu, \lambda) \sqrt{ 1 \over 2 \pi   \kappa \nu} 
e^{i \nu/\kappa }\, .   
\label{abcoef2}
\ea

Two crucial properties follow from eq. (\ref{abcoef2}):
firstly, the Planck distribution
characterizing the mean number of out-quanta in the 
in-vacuum (up to a normalisation factor, it
is obtained from $\vert \beta_{\nu, \lambda} \vert^2$),
and secondly, the existence of a constant flux of out-quanta.
The stationarity follows from the fact that the
phases of $\beta$ and $\alpha$ are both 
proportional to $\nu^{-i \la/\kappa}$.
This implies indeed that the 
value of the energy flux is constant.
To prove it, 
we recall that the renormalized value
of the energy flux in the in-vacuum,
when expressed in terms of the Bogoliubov coefficients, is
\ba
&&{dE\over du} =  \int_0^{\infty} d\la \int_0^{\infty} d\la^\p
{\sqrt{\la \la^\p} \over
2\pi} \left[ e^{-i( \la -\la^\p)u }\left( \int_0^\infty d\nu
\bt^*_{\nu, \la}
 \bt_{\nu, \la^\p}
\right)\right.
\nonumber\\
&&\quad\quad\quad\quad\quad\quad\quad\quad\quad
\quad\quad\quad\quad \left.
- \mbox{Re} \left( e^{i (\la + \la^\p)u}
\int_0^\infty d\nu \al_{\nu, \la} \bt_{\nu, \la^\p} \right) \right]\,
.
\label{Tuuq}
\ea
By using eq. (\ref{abcoef2}) and performing the integral over $\nu$,
 one immediately obtains that the 
second term, the interfering one, vanishes and that the 
first term is constant.

We shall now determine how these properties are 
affected by the fluctuations of the geometry
described by eq. (\ref{2.1}).
Under our restriction to s-waves and 
neglection of the potential barrier, 
it suffices to repeat the same procedure
with the modified function $V(u)$, as if we were
considering 2D propagation.
The asymptotic expression we shall use is given by
\be
 -\kappa_r V(u) = \tilde{V}_0 +  e^{-\kappa_r u} [ 1 +  
A_1 \sin(\om u + \varphi_1) +A_2 \sin(2 \om u + \varphi_2)]\, ,
\label{asV}
\ee
where the values of $\tilde{V}_0$, $A_1$, and $A_2$ can be found
in eqs. (\ref{4.49}), (\ref{5.4}), and (\ref{5.3d}).
We have kept the constant shift in $V$ since it will 
modify the absolute phase of the Bogoliubov coefficients.
The symbol $\kappa_r$ designates the `renormalized'
surface gravity introduced in eq. (\ref{5.10}). 
We have indeed absorbed the linear term in $u$
of eq. (\ref{5.3}) in the redefinition of $\kappa$. 
As before, since $\kappa_r - \kappa=
O(\mu_0^2)$ and since we shall work up to quadratic 
corrections in $\mu_0$, the modification of the terms 
proportional to $A_1$ and $A_2$ are irrelevant.

 To reveal the nature of the 
modifications induced by the time dependent fluctuations
of the metric, it is appropriate to analyse the
$\beta$ coefficient.  
Up to quadratic order in the metric fluctuations
of amplitude $\mu_0$, the modified $\beta$ coefficient
reads
\ba
\beta^{mod}_{\nu, \lambda}
&=& \int^\infty_{-\infty} du {e^{-i  \lambda  u} \over \sqrt{ \pi
\lambda^{-1}} }
{e^{-i \nu  V(u)} \over \sqrt{4 \pi \nu} } 
\nonumber\\
&=& \int du {e^{-i\lambda  u} \over \sqrt{ \pi \lambda^{-1}} }
{e^{i  \nu (\tilde{V}_0 + e^{-\kappa_r u})/\kappa_r } \over \sqrt{4 \pi
\nu} } 
\left[ 1 - { 1 \over 2} ({ \nu \over \kappa_r })^2 e^{-2 \kappa_r u} 
A^2_1 \sin^2(\om u + \varphi_1) \right.
\nonumber\\
&& \quad\quad
\quad\quad\quad\left.
+ i { \nu \over \kappa_r} e^{-\kappa_r u}
\{
A_1 \sin(\om u + \varphi_1) +A_2 \sin(2 \om  u + \varphi_2)
\}  \right]\, .
\ea

By decomposing the sines
into imaginary exponentials and by using several times
 the relation $\Gamma(x+1) = x \Gamma(x)$ we get
\ba
\beta^{mod}_{\nu, \lambda} &=& \sqrt{1 \over 2\pi  \kappa_r \nu}  
 e^{i\nu   \tilde{V}_0/\kappa_r} \times \left\{
B_r(\nu, \la) \left[ 1 - { A_1^2 \over 4 \kappa_r^2}
(\la -i \kappa_r ) \la \right]
\right.
\nonumber
\\
&&
  -  \sqrt{\la}{ A_1  \over 2 \kappa_r} 
\left[ 
B_r(\nu, \la - \om)  \sqrt{\la - \om} \; e^{i \varphi_1} 
- B_r(\nu, \la + \om) \sqrt{\la + \om} \; e^{-i \varphi_1} 
\right] 
 \nonumber
\\
&& 
-  \sqrt{\la}{A_2   \over 2 \kappa_r}   \left[ 
B_r(\nu, \la - 2\om) \sqrt{\la - 2\om} \; e^{i \varphi_2} 
- 
B_r(\nu, \la + 2\om) \sqrt{\la + 2\om} \; e^{-i \varphi_2} \right]
\nonumber
\\
&& 
+   \sqrt{\la}{ A_1^2 \over 8 \kappa_r^2 } \left[ B_r(\nu, \la - 2\om) \ 
({ \la - 2\om} - i \kappa_r)  \sqrt{\la - 2\om}  
 \; e^{2i \varphi_1}
\right.
 \nonumber
\\ 
&& \left. \left. \quad \quad \quad \quad 
+ B_r(\nu, \la + 2\om) 
({ \la + 2\om } - i \kappa_r )  \sqrt{\la + 2\om}  
\; e^{-2i \varphi_1}
\right] \right\}\, .
\label{modB}
\ea
Here $ B_r(\nu, \la)$ designates the function $B(\nu, \la)$
defined in eq. (\ref{Bd}) with the renormalized surface gravity.

The replacement of $B(\nu, \la)$ by $B(\nu, \la \pm \om)$  in 
the last three terms of (\ref{modB})
indicates that $\om$, the frequency of the fluctuating metric, 
enters into the expressions in such a way that the `effective'
frequency, i.e. the one which weighs the new amplitudes,
is $\la \pm \om$.

Physically, this leads to a modification of the mean
(quantum averaged and time averaged) number of 
 quanta which reach ${\cal J}^+$ per unit $u$ time. This 
averaged flux is
\be 
\bk{\bar n_\la} =  { \int^N d\nu \vert \beta^{mod}_{\nu, \la}
\vert^2 \over \int^N d\nu/\kappa_ r \nu }
\label{prat}
\ee
We have implemented time average
by integrating over $\nu$ up to the cut-off frequency $N$ and dividing
the resulting expression by $\int^N  d\nu/\kappa_r \nu = \Delta u$.
This last equality follows from the resonance condition, eq. (\ref{resc}),
and its validity requires that $\Delta u \gg 1/\kappa_r $, as 
in usual Golden Rule estimates.

In this ratio, to quadratic order in $\mu_0$,
only four terms contribute.
The first two terms of the first line of eq. (\ref{modB})
and those resulting from the square of the second line.
This is because the integral of all other terms 
are crossed terms
of the form $B(\nu, \la) B^*(\nu, \la  \pm \om)$
which lead to oscillatory integrands containing
$\; e^{ \pm i\om u_{sp}(\nu)}$.
These terms do not contribute to the production
rate, eq. (\ref{prat}), in the limit of large $\kappa \Delta u$.
 
By combining the four non-vanishing contributions we get 
\ba\n{6.11}
\bk{\bar n_\la} &=& {1 \over 2 \pi} \left\{
{ 1 \over e^{2 \pi \la / \kappa_r  } - 1}
\left( 1 -  ({A_1\over \kappa_r})^2 { \la ^2  \over 2 }
\right)
\right.
\nonumber\\
&&\left. \quad \quad
 + ({ A_1  \over 2 \kappa_r })^2 {\la} \left[ 
 { \la -\om\over e^{2 \pi (\la -\om) / \kappa_r  } - 1 } +
 { \la + \om \over e^{2 \pi (\la + \om) / \kappa_r  } - 1 }
\right] \right\} \quad \quad
\label{modF}
\ea

\begin{figure}[tb]
\centerline{\epsfig{file=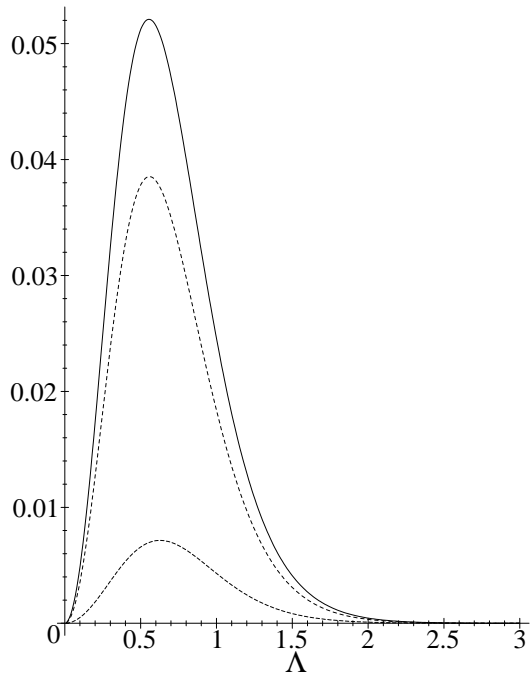, height=11.5cm}}
\caption[fig.2]{Function $F(\Lambda, \Omega)$ for different values of
dimensionless frequency $\Omega$: $\Omega=0.01$ -- solid line, 
$\Omega=0.1$ -- dashed line, and
$\Omega=0.5$ -- dotted line.}
\label{fig.2}
\end{figure}

Having obtained the spectrum of the particle number flux,
the spectrum of the energy flux is
\be\n{6.14}
{dE\over du\, d\lambda}=\lambda\, \bk{\bar n_\la}=
{\kappa_r \over 2\pi} \,\left[ f(\Lambda)
+{\mu_0^2 \over 2} F(\Lambda;\Omega)\right]\, ,
\ee
where $\Lambda=\lambda/\kappa_r$, $\Omega=\omega/\kappa_r$,
\be\n{6.15}
f(\Lambda)={\Lambda\over{\exp(2\pi\Lambda)-1}}\, ,
\ee
and
\be\n{6.16}
F(\Lambda;\Omega)={\pi\Lambda^2\over \Omega^2(1+\Omega^2)}\, f(\Omega)\,
\left[ f(\Lambda -\Omega)+f(\Lambda
+\Omega)- 2f(\Lambda)\right]\, .
\ee
Plots of the function $F(\Lambda;\Omega)$ for different values of
$\Omega$ are given in Figure~2. 

In order to obtain the constant part of the 
density of energy flux one must 
integrate (\ref{6.14}) over the frequency $\lambda$
\be\n{6.17}
{dE\over du}={\kappa_r^2\over 2\pi}\int_0^{\infty}
d\Lambda\,  \left[ f(\Lambda)
+{\mu_0^2 \over 2} F(\Lambda;\Omega)\right]\, .
\ee
Using formulas (\ref{B.1}) and (\ref{B.10}) from appendix B, it is
easy to verify that (\ref{6.17}) coincides with the expression
(\ref{5.11}) for $(dE/du)^{\ind{perm}}$ we obtained earlier.

The other physical consequence of the corrections  terms to $\alpha$
and $\beta$ is the following. The mean instantaneous energy flux now
oscillates around its time average with the harmonics of $\om$. This
can be seens from the oscillatory terms  $B(\nu, \la) B^*(\nu, \la 
\pm (2) \om)$ which behave in $\; e^{ \pm i (2) \om u_{sp}(\nu)}$
when parametrized by $\kappa u_{sp} = \ln \nu/ \la$, the location of
the dominant contribution to $\alpha_{\nu, \la}$. The amplitudes of
the fluctuations are linear in $A_1$ and $A_2$. Moreover, the phase
shifts of the modified coefficients with respect the unperturbed ones
differ for $\alpha$ and $\beta$. This implies that the second term in
eq. (\ref{Tuuq}) no longer vanishes even though it still does not
contribute to  the mean (time average) energy flux.

\section{Conclusion} 

In this paper we studied how the fluctuations of the geometry
near the black-hole horizon affect Hawking radiation. 
To characterize the fluctuations of geometry, we used York's model
\cite{York:83} in which they are described by an
in-coming Vaydia metric with a time dependent mass $m(v)$
which fluctuates around its mean with amplitude $\mu_0$
and period $2 \pi/\omega$.
For further simplicity, we only considered the $s$-modes 
of a quantum scalar massless
field and we also neglected the scattering 
by the gravitational potential barrier. 
Using these simplifying assumptions,
we reached the following conclusions.

First, the expectation value of the outgoing flux of energy is no 
longer constant. It now fluctuates around its time averaged value
with frequencies given by harmonics of $\om$ and with amplitudes
starting with a term linear in $\mu_0$, see eq. (\ref{5.8}). 
The fact that the
phase of the  fluctuations of the expectation value of $dE/du$  is
well defined results from the fact that the fluctuations  we
considered were treated classically. In a more quantum mechanical
treatment, these well defined phases will probably be replaced by a
more diffuse  ensemble of phases.

Secondly, the time averaged value of the outgoing flux of energy is
modified. One part of this modification is connected with  the
renormalization of the surface gravity of the fluctuating black hole
given by expression (\ref{5.10}). The other part 
is an additional factor given by 
$\mu_0^2\pi\Omega/(\exp(2\pi\Omega)-1)$, see eq. (\ref{5.11}).  Both
changes are second order in $\mu_0$.
More surprisingly, they decrease for large $\Omega$.
Indeed, one might have feared that fast metric fluctuations
would lead to copious pair creation
\footnote{We are grateful to T. Jacobson for pointing out
this fact to us.}. We conjecture that we do not
find abundant production since we worked
in a 2 dimensional model in which the time dependence of the
metric does not affect directly massless modes, thanks to conformal
invariance.

Thirdly, the asymptotic spectrum of Hawking radiation  is also
modified. Besides the renormalization of the surface gravity which
shifts the temperature, the modified spectrum (\ref{modF})  contains
three additional correction terms.  The two last terms in that
equation contain  Bose thermal factors of the form
$1/(\exp(2\pi(\lambda\pm \omega)/\kappa)-1)$. In these relations, the
frequency of geometry fluctuations, $\pm \omega$,  plays the role of
a chemical potential.   The presence of such chemical potential is
reminiscent to superradiance.  This fact supports the general ideas
proposed by York since the appearance of these factors might be
expected  from the existence of a {\em quantum ergoshpere}.  Indeed,
due to quantum fluctuations,  the average position of the event
horizon is  moved by a term proportional to the second power $\mu_0^2$
of the
amplitude of  fluctuations, while the temporal position of the
apparent  horizon is fluctuating with  amplitude $\mu_0$.  An alternative
way to describe these fluctuations is to say that there exists a
blurring of the physical null cone at the unperturbed horizon.
Because of the existence of negative energy states 
inside the unperturbed black hole matter can escape from the narrow
region close to the horizon. This leakage of energy is seen as
Hawking radiation \cite{York:83}. Under the same conditions one can
expect an additional amplification of Hawking quanta while they are
propagating close to the fluctuating horizon. The amplification factor we
got in the expression for the modified spectrum of Hawking radiation
may be considered as an indication to this effect.

The modifications in the black hole temperature and surface
area in the presence of metric fluctuations raise the question
about the modifications of black hole thermodynamics.
If we identify the energy of the
system $E$ with the averaged mass of the black hole $M$, and the
temperature of the black hole, $T$, with $\kappa_r/2\pi$, then the first
law $dE=T\, dS$ and eqs.(\ref{3.10a}) and (\ref{5.10})
define the averaged entropy to be
\be\n{7.2}
\bar S={ \bar {\cal A}\over 4}
[1+{\mu_0^2\over 2(1+\Omega^2)}]^{-2}\sim
{ \bar {\cal A}\over 4}\left(1-{\mu_0^2\over 1+\Omega^2}\right)\, .
\ee

Therefore, one looses the relationship between 
the entropy and a fourth of the area.

If one writes the amplitude of the fluctuations as 
$\mu_0 =\alpha m_{\ind{Planck}}/M$ 
where $\alpha$ is dimensionless,
one has
\be\n{7.3}
\bar S=4\pi \, {M^2\over m_{\ind{Planck}}^2} -s\, ,
\ee
where
\be\n{7.4}
s={4\pi \, \alpha^2\over 1+\Omega^2}
\ee
does not depend on the black hole mass.  It
is worth noticing that this modification 
of the black hole entropy
is exactly of the same form as in theories with corrections
quadratic in the curvature \cite{IyWa:94}. 
We recall that these corrections arise in the effective action 
by taking, in the one-loop approximation,
 the average of the gravitational equations 
over quantum fluctuations of the metric.
This observation might give a possible explanation to the
origin of the similarity between (\ref{7.3}) and the results of Ref.
\cite{IyWa:94}. Similarily,
it would be interesting to find the relation 
between the results obtained in this
paper and the quantum treatments of
\cite{Miko:97}--\cite{BuRaMi:97}.

Even though these results were obtained in
an extremely simplified model in which the metric fluctuations
were treated classically, we believe that they 
indicate what might be the impact of the quantum fluctuations 
of the near horizon 
geometry on black hole radiance.

\bigskip

\vspace{12pt}
{\bf Acknowledgements}:\ \ The authors thank Slava Mukhanov for
discussions. This work was  supported  by   NATO Grant CRG.972079.
One of the authors (V.F.) is grateful to the Natural Sciences and
Engineering Research Council of Canada and to the Killam Trust for
their financial support.

\newpage

\appendix

\section{Calculation of Function $S(x)$.}\label{sA.1}
\setcounter{equation}0

To calculate function 
\[
S(x) = \int_x^{\infty}\frac{d\xi \,e^{i\Omega\xi}}{\xi^{1-i\Omega}}\,
\int_{\xi}^{\infty}\frac{d\eta \,e^{-i\Omega\eta}}{\eta^{1+i\Omega}}
\]
\be
\hspace{.95cm} = (i\Omega )^{i\Omega}\,\int_x^{\infty}
\frac{d\xi \,e^{i\Omega
\xi}}{\xi^{1-i\Omega}}\,\Gamma (-i\Omega ,i\Omega\xi )\,, \n{A.1}
\ee
we use the following general result, which can be found in
\cite{PrBrMa:86} 
(volume II relation 2.10.3.6)
\[
\int_a^{\infty}x^{\alpha -1}\,(x-a)^{\beta -1}\,e^{cx}\,\Gamma(\nu,cx)\,dx
\]
\[
= a^{\alpha + \beta -1}\,\Gamma(\nu )\,B(\beta ,1-\alpha -\beta )\,{}_1F_1
(\alpha;\alpha +\beta ; ac)
\]
\[
-\,\frac{ a^{\alpha + \beta +\nu -1}c^{\nu}}{\nu} 
\,B(\beta ,1-\alpha -\beta -\nu )\,{}_2F_2
(\alpha +\nu ,1;\nu +1,\alpha +\beta +\nu ; ac)
\]
\be
-\,\frac{\pi c^{1-\alpha - \beta}}{\sin [(\alpha +\beta +\nu )\pi 
]}\,\frac{\Gamma 
(\alpha +\beta -1)}{\Gamma(1-\nu)} \,{}_1F_1(1-\beta ;2-\alpha -\beta ; 
ac)\,.
\n{A.2}
\ee
Here
\be
B(x,y) = \frac{\Gamma (x)\,\Gamma (y)}{\Gamma(x+y)}\,. \n{A.3}
\ee
This relation is valid for $a>0$,  $\mbox{Re}\,\beta >0$, 
$\mbox{Re}\,(\alpha +\beta +\nu )<2$; $|\,\mbox{arg}\,c\,|<\pi$. 
For a particular case 
$\beta = 1$ one has
\[
\int_a^{\infty}x^{\alpha -1}\,e^{cx}\,\Gamma(\nu,cx)\,dx
= - \frac{a^{\alpha }\,\Gamma (\nu )}{\alpha }\,{}_1F_1(\alpha;1+\alpha ; ac)
\]
\[
+\,\frac{ a^{\alpha +\nu }\,c^{\nu}}{\nu (\alpha +\nu )} \,{}_2F_2
(\alpha +\nu ,1;1+\nu ,1+\alpha +\nu ; ac)
\]
\be
+\,\frac{\pi c^{-\alpha }}{\sin [(\alpha +\nu )\pi ]}\,\frac{\Gamma 
(\alpha )}{\Gamma(1-\nu)}\,. \n{A.4}
\ee 
The integral in (\ref{A.1}) is of the form (\ref{A.4}) with
\be
\alpha = i\Omega \,, \quad \nu = -i\Omega \,, \quad a = x\,, \quad 
c = i\Omega \,. \n{A.5}
\ee
For this case $\alpha +\nu  = 0$, and the last two terms in (\ref{A.4}) 
are 
divergent. To consider this limit, we define $\gamma = \alpha +\nu $ and 
rewrite the sum of these two divergent terms in the following form
\[
\frac{ a^{\alpha +\nu }\,c^{\nu}}{\alpha (\alpha +\nu )} \,{}_2F_2
(\alpha +\nu ,1;1+\nu ,1+\alpha +\nu ; ac)
+\,\frac{\pi c^{-\alpha }}{\sin [(\alpha +\nu )\pi ]}\,\frac{\Gamma 
(\alpha )}{\Gamma(1-\nu)}
\]
\be
= \frac{c^{\nu -\gamma }}{\nu}\,Z\,, \n{A.6}
\ee
where
\be
Z = \frac{1}{\gamma }\left[y^{\gamma}\,{}_2F_2(\gamma ,1;1+\nu ,1+\gamma ;
y) + \frac{\pi\gamma\nu}{\sin (\gamma\pi )}\,\frac{\Gamma (\gamma -\nu )}
{\Gamma (1-\nu )}\right], \n{A.7}
\ee
and $y=ac$.

Using the expression for the hypergeometric function
\be
{}_2F_2(\gamma ,1;1+\nu ,1+\gamma ;y) = 1  
+ \gamma\sum_{n=1}^{\infty}\frac{y^n}{(\gamma + n)(1+\nu )_n}\,, \n{A.8}
\ee
where $(1+\nu )_n = \Gamma (1+ n +\nu )/\Gamma (1+ \nu)$, we can write 
for 
small $\gamma$ 
\be
{}_2F_2(\gamma ,1;1+\nu ,1+\gamma ;y) \approx 1  
+ \gamma f(\nu ;y)\,, \n{A.9}
\ee
where
\be
f(\nu ;y) =\sum_{n=1}^{\infty}\frac{y^n}{n(1+\nu )_n }\,. \n{A.10}
\ee
We also have
\be
\frac{\Gamma (\gamma -\nu )}{\Gamma (1-\nu )} \approx - 
\frac{1}{\nu } + \frac{\Gamma '(-\nu )}{\Gamma (1-\nu )}\,\gamma = 
- \frac{1}{\nu } - \frac{1}{\nu }\,\psi (-\nu )\,\gamma\,, \n{A.11}
\ee
where $\psi (z) = d\ln \Gamma (z)/dz$.

Substituting expression (\ref{A.9}) and  (\ref{A.11}) into  (\ref{A.7}) 
and 
expanding $\sin (\gamma\pi ) = \gamma\pi (1-\frac{1}{6}\gamma^2\pi^2 + 
\ldots)$, we get
\[
Z = \frac{1}{\gamma}\left[y^{\gamma}(1+\gamma f(\nu ,y)) -1 - \gamma
\psi (-\nu )\right]
\]
\be 
\hspace{.5cm} = \frac{y^{\gamma}-1}{\gamma} + f(\nu ,y) - \psi (-\nu ) 
\approx 
\ln y + f(\nu ,y) - \psi (-\nu )\,. \n{A.12}
\ee
Thus for $\nu = -\alpha$ we have
\[
\int_a^{\infty}x^{\alpha -1}\,e^{cx}\,\Gamma(-\alpha ,cx)\,dx
= -\frac{a^{\alpha }\,\Gamma (-\alpha )}{\alpha 
}\,{}_1F_1(\alpha;1+\alpha ; ac)
\]
\be
\hspace{5.1cm} -\,\frac{c^{-\alpha}}{\alpha }\,[\ln (ac) + f(-\alpha ;ac) 
- 
\psi (\alpha )]\,. \n{A.13}
\ee 
Using this result, we obtain for $S(x)$ the following expression
\[
S(x) = \frac{i}{\Omega }\,e^{i\Omega\ln \Omega + i\Omega \ln
x}\,e^{-\pi\Omega/2}\, \Gamma
(- i\Omega )\,{}_1F_1( i\Omega;1+ i\Omega; i\Omega x) 
\]
\be
\hspace{1.0cm} +\,\frac{i}{\Omega}\,
[\ln (i\Omega x) + f(-i\Omega ; i\Omega x) - \psi ( i\Omega )]\,. \n{A.14}
\ee

\section{Calculation of Integrals.}\label{sB.1}
\setcounter{equation}0

In this appendix we demonstrate that
\be\n{B.1}
J\equiv \int_0^{\infty}{d\lambda\, \lambda^2\,
(\lambda+\Omega)\over{\exp[2\pi(\lambda+\Omega)]-1}}+
\int_0^{\infty}{d\lambda\, \lambda^2\,
(\lambda-\Omega)\over{\exp[2\pi(\lambda -\Omega)]-1}}=
{1\over 120}(10\Omega^4+10\Omega^2+1)\, .
\ee

First we notice that
\be\n{B.2}
J=\lim_{p\rightarrow 0}\, \left[J_+(p)+J_-(p)\right]\, ,
\ee
where
\be\n{B.3}
J_{\pm}(p)=\int_0^{\infty}{d\lambda\, \lambda^2\,
(\lambda\pm\Omega)\exp(-p|\lambda|)\over{\exp[2\pi(\lambda\pm\Omega)]-1}}\,
.
\ee
Making change of variable of integration $\lambda\rightarrow -\lambda$
in the expression for $J_-(p)$ we get
\be\n{B.4}
J_-(p)=\int_{-\infty}^0 d\lambda\,\lambda^2\,
(\lambda+\Omega)\exp(p\lambda)+
\int_{-\infty}^0 {d\lambda\, \lambda^2\,
(\lambda+\Omega)\exp(-p|\lambda|)\over{\exp[2\pi(\lambda+\Omega)]-1}}\,
.
\ee
Thus we have
\be\n{B.5}
J_+(p)+J_-(p)=\int_{-\infty}^0 d\lambda\,\lambda^2\,
(\lambda+\Omega)\exp(p\lambda)+
\int_{-\infty}^{\infty} {dx\, (x-\Omega)^2\,
x\exp(-p|x-\Omega|)\over{\exp(2\pi x)-1}}\,
.
\ee
In the last integral in the right hand side we made change of
variables $\lambda=x-\Omega$.

Consider now the integral
\be\n{B.6}
\int_{-\infty}^{0} {dx\, (x-\Omega)^2\,
x\exp(-p|x-\Omega|)\over{\exp(2\pi x)-1}}\,.
\ee
By changing variables $x\rightarrow -x$ it can be identically rewritten as
\be\n{B.7}
\exp(-p\Omega)\int_0^{\infty} dx \, x \, (x+\Omega)^2\, \exp(-px)+
\exp(-p\Omega)\int_0^{\infty}{dx \, x \, (x+\Omega)^2\,
\exp(-px)\over{\exp(2\pi x)-1}}\, .
\ee
Using (\ref{B.5}) and (\ref{B.7}) and taking the limit $p\rightarrow
0$ we get
\be\n{B.8}
J=2\int_0^{\infty}{dx \, x \, (x^2+\Omega^2)\over{\exp(2\pi
x)-1}}+\Delta J\, ,
\ee
where
\[
\Delta J=\lim_{p\rightarrow 0}\left[\int_{-\infty}^0 dx\, x^2\,
(x+\Omega)\exp(px)
+\exp(-p\Omega)\, \int_0^{\infty} dx\, x\,
(x+\Omega)^2\exp(-px)\right]
\]
\be\n{B.9}
=\lim_{p\rightarrow 0}\left[\left(-{6\over p^4}+{2\Omega\over p^3}
\right)+e^{-p\Omega}\left({6\over p^4}+{4\Omega\over p^3}
+{\Omega^2\over p^2}\right)\right] = {1\over 12}\Omega^4\, .
\ee
The integrals which enter (\ref{B.8}) can be easily calculated
\be\n{B.10}
\int_0^{\infty}{dx \, x \over{\exp(2\pi
x)-1}}={1\over 24}\, ,
\hspace{0.5cm}
\int_0^{\infty}{dx \, x^3 \over{\exp(2\pi
x)-1}}={1\over 240}\, .
\ee
Combining these results we get (\ref{B.1}).


\newpage

\begin{thebibliography}{000}

\bibitem{Hawk:75} S.W. Hawking, Comm. Math.
Phys. {\bf 43} (1975) 199.

\bibitem{BD} N.D. Birrel and P.C.W. Davies, 
{\it Quantum Fields in Curved Space},
Cambridge University Press, 1982.

\bibitem{FrNo:98} V. Frolov and I. Novikov. {\it Black Hole Physics:
Basic Concepts and New Developments,} Kluwer Academic Publ., 1998.


\bibitem{tHooft} C. R. Stephens, G. 't Hooft and B. F. Whiting,
Class. Quant. Grav. {\bf 11} (1994) 621.

\bibitem{CoJa} T. Jacobson, Phys. Rev. D {\bf 44} (1991) 1731,
Phys. Rev. D {\bf 48} (1993) 728.

\bibitem{Parent} F. Englert, S. Massar and R. Parentani,
Class. Quantum Grav. {\bf 11} (1994) 2919.    

\bibitem{VV} Y. Kiem, H. Verlinde, E. Verlinde, Phys. Rev. D {\bf 52}
(1995) 7053.

         
\bibitem{Unru:95} W. G. Unruh, Phys. Rev. D {\bf 51} (1995) 2827.

\bibitem{Parent1}  R. Brout, S. Massar, R. Parentani, 
and Ph. Spindel, Phys. Rev. D {\bf 52}
(1995) 4559.

\bibitem{CoJa:96} S. Corley and T. Jacobson, Phys. Rev. D {\bf 54} 
(1996) 1568.



\bibitem{KrWi} P. Kraus and F. Wilczek, Nucl. Phys. B {\bf 433} (1995)
403.

\bibitem{KKV} E. Keski-Vakkuri and P. Kraus, Nucl. Phys. B {\bf 491} 
(1997) 249.

\bibitem{Schw:61} J. Schwinger, J. Math. Phys. {\bf 2} (1961) 407.

\bibitem{Keld:64} L. V. Keldysh, Zh. Eksp. Teor. Fiz. {\bf 47} (1964)
1515; [Engl. Trans. Sov. Phys. JEPT {\bf 20} (1965) 1018.]

\bibitem{FeVe:63} R. Feynman and F. Vernon, Ann. Phys. (NY) {\bf 24}
(1963) 118.

\bibitem{FeHi:65} R. Feynman and A. Hibbs, {\em Quantum Mechanics and
Path Integrals}, McGraw-Hill, New York, 1965.


\bibitem{Hu:99} B. L. Hu. Preprint   gr-qc/9902064.

\bibitem{CaHu:98} A. Campos and B. L. Hu. Preprint gr-qc/9812034.

\bibitem{MaVe:98a} R. Mart\'{i}n and E. Verdaguer. Preprint
gr-qc/9811070.

\bibitem{MaVe:98b} R. Mart\'{i}n and E. Verdaguer. Preprint
gr-qc/9812063.


\bibitem{York:83} J. W. York, Jr., Phys. Rev. D {\bf 28} (1983)  2929.

\bibitem{MTW:73} C.W. Misner, K.S. Thorne, and J.A. Wheeler,
{\it Gravitation}, San Francisco: Freeman, 1973.

\bibitem{HuSh:97} B. L. Hu and K. Shiokawa, Phys. Rev. D {\bf 57} (1998)
3474.

\bibitem{Bardeen} J. M. Bardeen, Phys. Rev. Lett. {\bf 46} (1981) 382.

\bibitem{Massar} S. Massar, Phys. Rev. D {\bf 52} (1995) 5857.



\bibitem{PrBrMa:86} A. P. Prudnikov, Yu. A. Brychkov, and O. I.
Marichev. {\it Integrals and Series}, Gordon and Beach Science Publ.,
v.II, p.150, 1986.

\bibitem{IyWa:94} V. Iyer and R. Wald, Phys. Rev. D {\bf 50} (1994) 846.
\bibitem{MiRa:97} A. Mikovi\'{c} and V. Radovanovi\'{c}, Class. Quantum
Grav. {\bf 14} (1997) 2647.
\bibitem{Miko:97} A. Mikovi\'{c}, Phys. Rev. D {\bf 56} (1997) 6067.
\bibitem{CrMiNa:97} J. Cruz, A. Mikovi\'{c}, and J. Navarro-Salas, Phys.
Lett. B {\bf 395} (1997) 184.
\bibitem{BuRaMi:97} M. Buri\'{c}, V. Radovanovi\'{c}, and A.
Mikovi\'{c}. Preprint gr-qc/9804083. 

\end{thebibliography}
\end{document}